\documentclass[letterpaper,10pt]{article}
\usepackage{osameet2}

\usepackage{amsmath,amssymb}
\usepackage[utf8]{inputenc}

\usepackage[dvips,colorlinks=true,bookmarks=false,citecolor=blue,urlcolor=blue]{hyperref} %latex w/dvips

\begin{document}

\title{Fish embryo multimodal imaging by laser Doppler digital holography}

\author{Nicolas Verrier$^1$, Daniel Alexandre$^1$, Pascal Picart$^{2,3}$, Michel Gross$^1$}
\address{$^1$Laboratoire Charles Coulomb - UMR 5221 CNRS-UM2 CC 026 Universit\'{e} Montpellier II Place Eug\^{e}ne
Bataillon 34095 Montpellier cedex, France\\
$^2$LUNAM Universit\'{e}, Universit\'{e} du Maine, CNRS UMR 6613, LAUM, Avenue Olivier Messiaen, 72085 Le Mans Cedex 9, France\\
$^3$\'{E}cole Nationale Sup\'{e}rieure d'Ing\'{e}nieurs du Mans, rue Aristote, 72085 Le Mans Cedex 9, France
}
\email{michel.gross@univ-montp2.fr}

\begin{abstract}
A laser Doppler imaging scheme combined to an upright microscope is proposed.  Quantitative Doppler imaging in both velocity norm and direction, as well as amplitude contrast of either zebrafish flesh or vasculature is demonstrated.
\end{abstract}


\textbf{Citation}

M. Gross, N. Verrier, and P. Picart, "Fish embryo multimodal imaging by laser Doppler digital holography," in Imaging and Applied Optics 2014, OSA Technical Digest (online) (Optical Society of America, 2014), paper JTu4A.7.
\url{http://www.opticsinfobase.org/abstract.cfm?URI=SRS-2014-JTu4A.7}

\textbf{Conference Paper}

Signal Recovery and Synthesis
Seattle, Washington United States
July 13-17, 2014
ISBN: 978-1-55752-308-2
Joint Poster Session (JTu4A)

\section{Introduction}

Combination of laser Doppler holography~\cite{AtlanGrossRSI2007} with transmission microscopy is proposed to analyze  blood in fish embryos .
We have adapted a laser Doppler holographic setup to a standard  bio-microscope by carrying the two beams  (illumination of the object and reference) with  optical fibers.
Multimodal acquisition and analysis of the data is made  by adjusting the reference versus illumination  frequency offset and the shape of the  filtering zone in the Fourier space.  Considering the same set of data, we have extracted amplitude contrast of the whole fish embryo, or only  the moving blood vessels. Images where the flow direction is coded in RGB color are obtained.

\section{Experiment}

\begin{figure}[htbp]
\centering
\includegraphics[width = 12 cm]{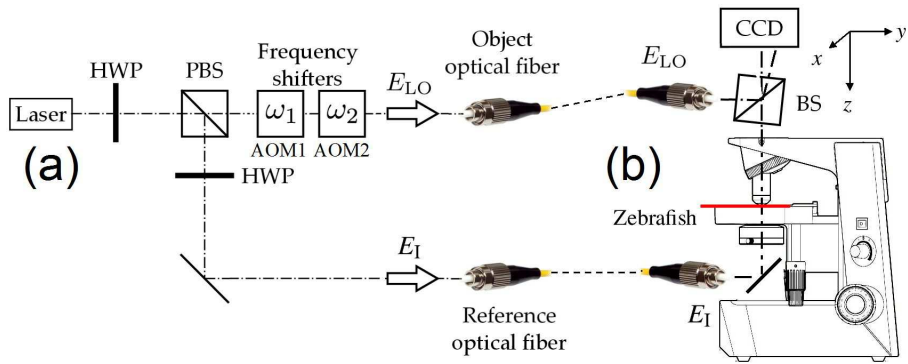}
\caption{Heterodyne digital holographic microscopy experimental arrangement (a) Injection part of the interferometer.  HWP: half wave plate; PBS: polarizing beam splitter; AOM1, AOM2: acousto optic modulators (Bragg cells). (b) Classical upright microscope used for the off-axis recombinaison of reference and object beams. BS: cube beam splitter; CCD: CCD camera.}\label{Setup}
\end{figure}
Our digital holography set-up consists in a Mach-Zehnder interferometer in which the recombining cube beam splitter is angularly tilted. The interferometer is split into an injection (a) and a recombination (b) part, which are connected, using optical fibers, to a classical upright microscope (Fig.\ref{Setup}).
The injection part  (Fig. \ref{Setup}(a)) is  used to build both object illumination (field $E_{\rm I}$) and reference (or local oscillator (LO) $E_{\rm LO}$) beams. Phase shifting interferometry and frequency scanning of the holographic detection frequency are made possible using the  heterodyne holography method \cite{LeClerc2000}.  The reference arm  is thus  frequency shifted by using two acousto-optic modulators AOM1 and AOM2, operating at a controlled angular frequency difference $\Delta\omega$. Reference and object fields are injected into two mono-mode optical fibers.
The object is imaged by a microscope objective (MO: NA= 0.25, G=10), mounted on an upright microscope (Fig. \ref{Setup}(b)).  The illumination field ($E_{\rm I}$) is scattered by the studied fish embryo (zebrafish) yielding the  signal field ($E$).  Signal ($E$) and reference ($E_{LO}$) fields are combined using an angularly tilted beam splitter cube (BS). Interferences (i.e. $E+ e^{j\Delta \omega t} E_{LO}$) between both fields are recorded on a 1360$\times$1024 pixel (6.45 $\mu$m square pitch) 12-bits CCD camera operating at $\omega_{\rm S} / (2\pi) \leq 10\ {\rm Hz}$. Recorded data are cropped to $1024\times 1024$ for FFT (Fast Fourier Transform) calculations. All the driving signals are synchronized by a common 10 MHz clock.

%\section{Amplitude contrast extraction}

\begin{figure}[htbp]
\centering
\includegraphics[width = 5.2cm]{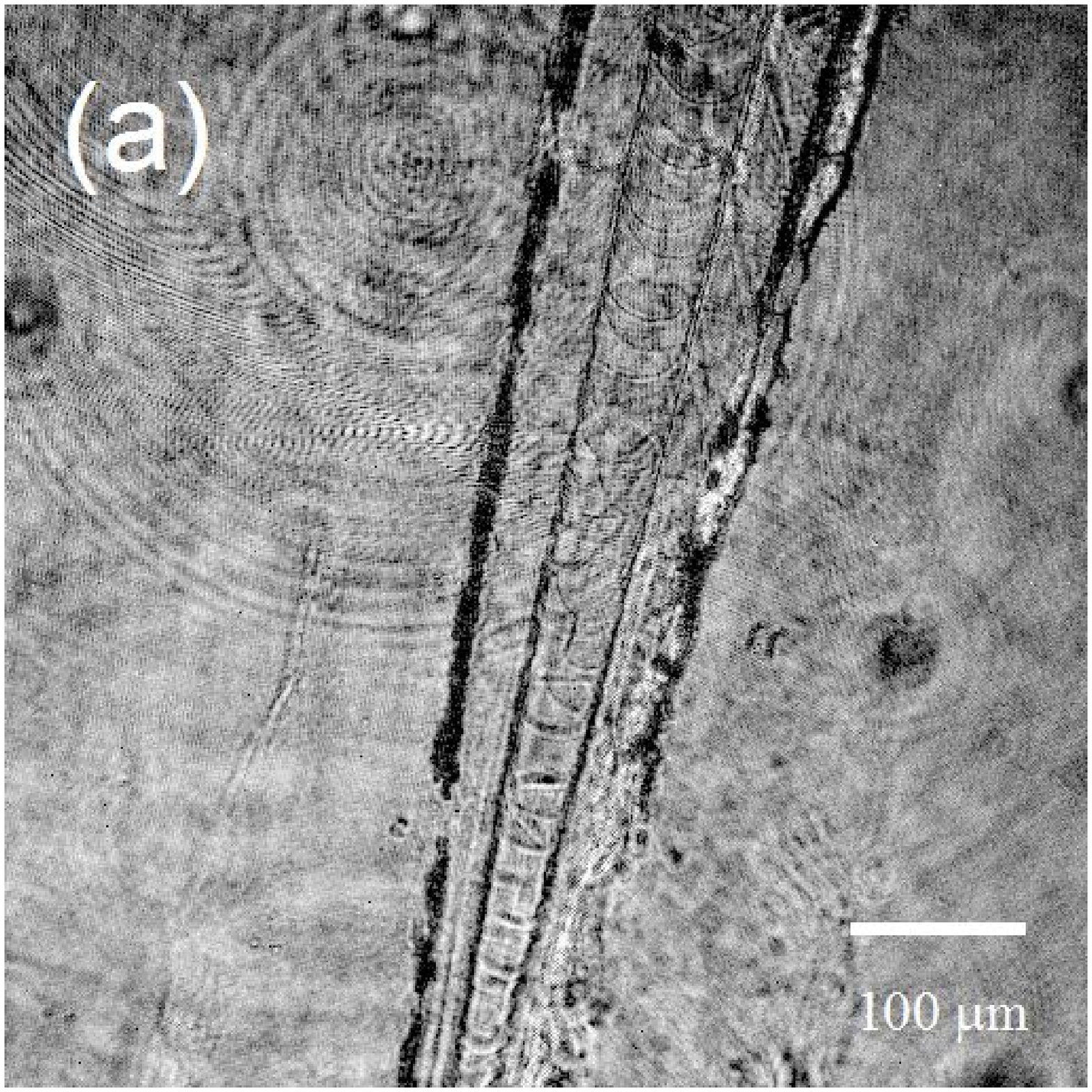}
\includegraphics[width = 5.2cm]{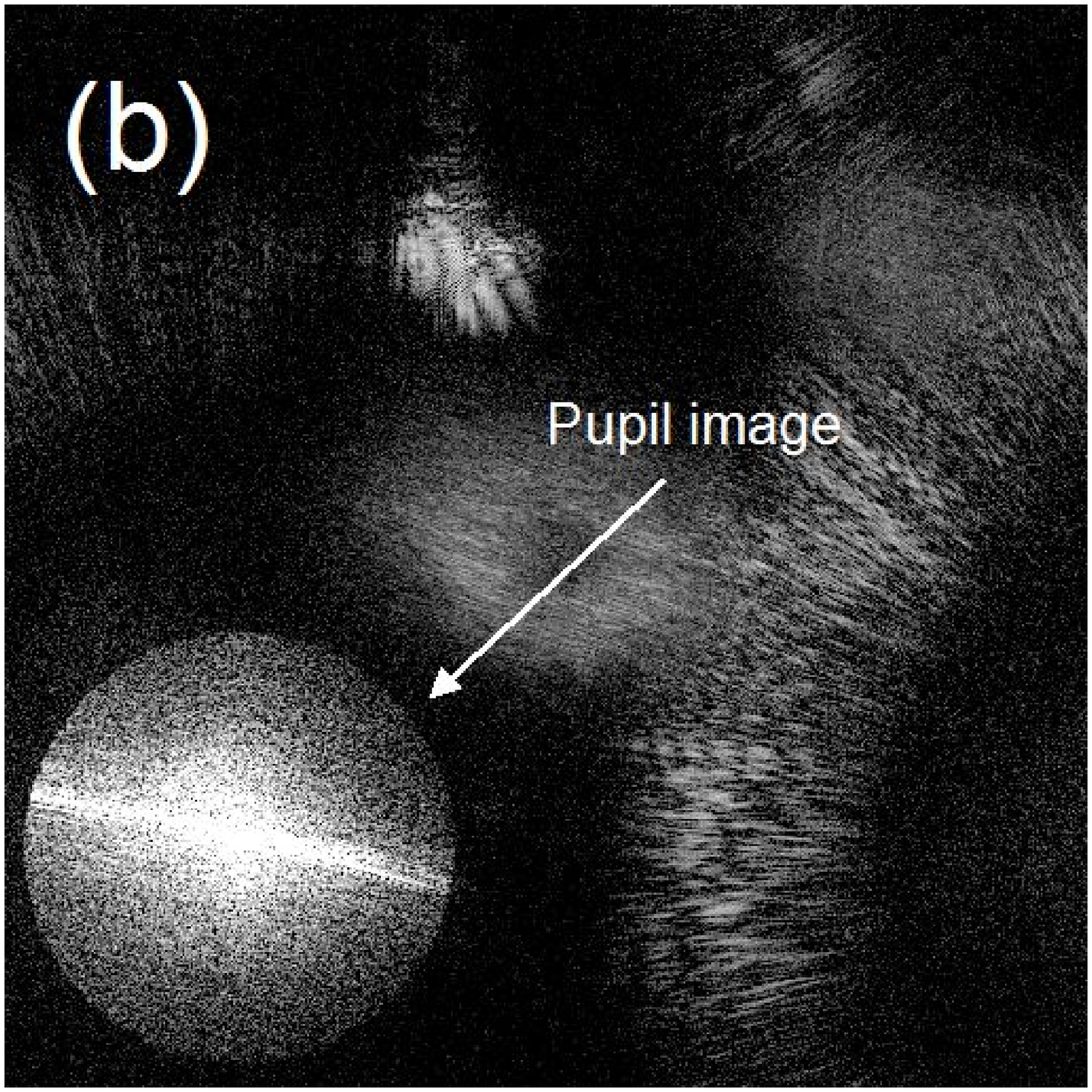}\\
\includegraphics[width = 5.25cm]{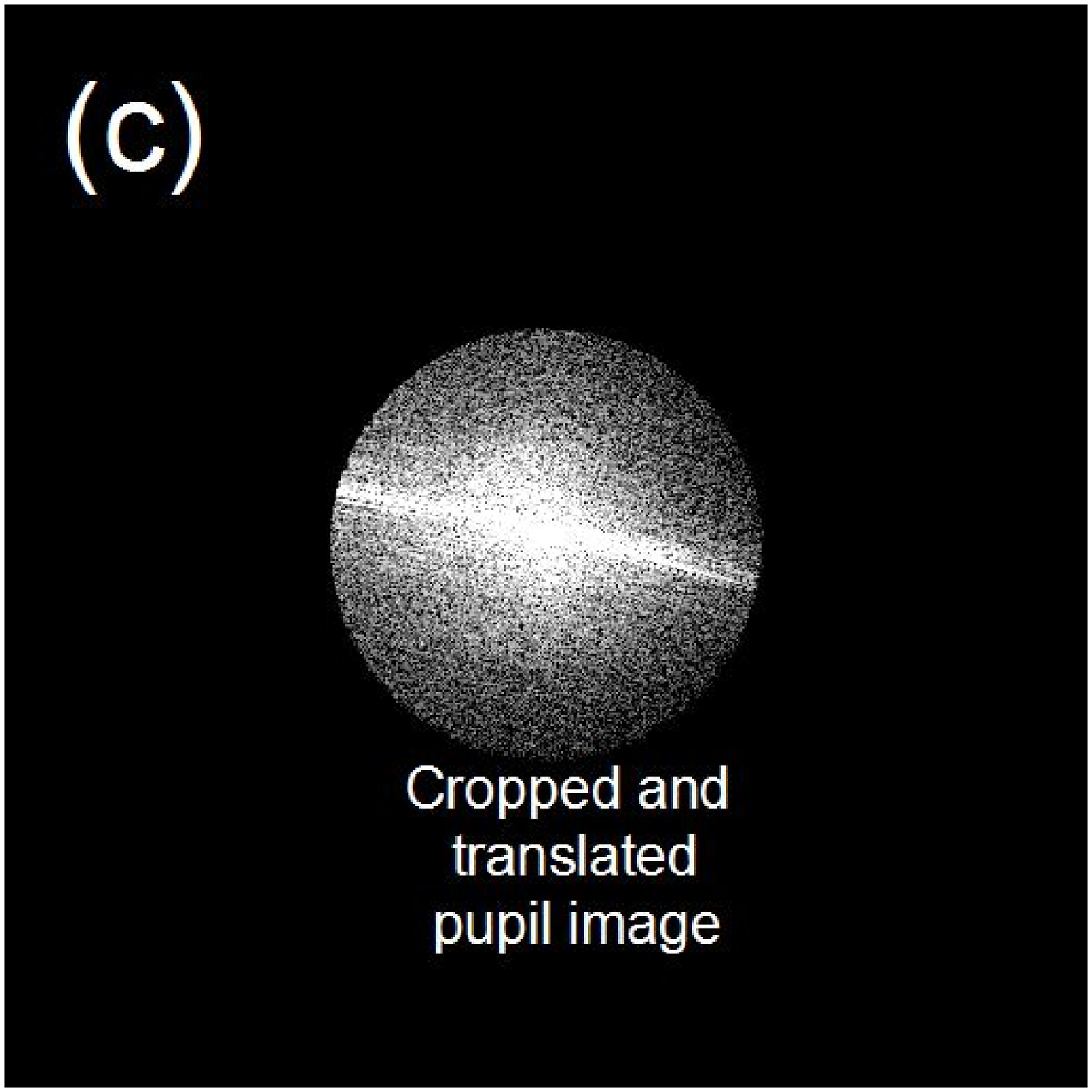}
\includegraphics[width = 5.2cm]{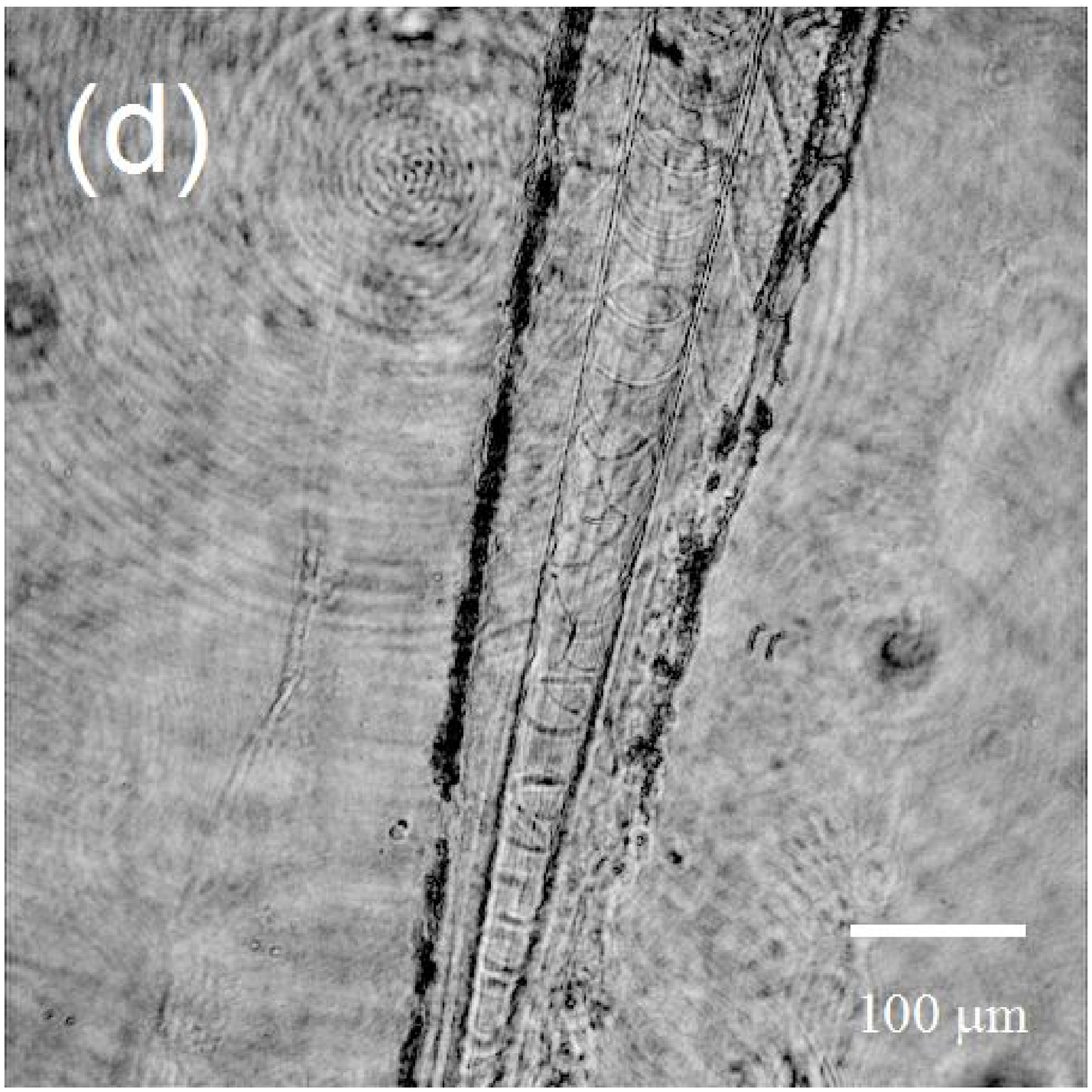}
\caption{Spatial filtering and reconstruction principle.  Holograms $H(x,y)$ (a), $H_1(k_x,k_y)$ (b), $H_2(k_x,k_y)$ (c) and $H_3(x,y,z=0)$ (d). The display is made in arbitrary Log scale for the average intensity  $\langle |H_{X}|^2\rangle$.}\label{fig_BW}
\end{figure}

Figure \ref{fig_BW} illustrates  our  reconstruction procedure \cite{warnasooriya2010imaging,verpillat2011dark} with   non-Doppler shifted images. We have considered  here four phase holograms: $H=(I_0-I_2)+j(I_1-I_3)$ (see Fig. \ref{fig_BW}(a)) where $I_0$..$I_3$ are 4 successive camera frames. $H$ is refocused in the objective pupil plane leading to $H_1\left(k_x,k_y\right)=\mathcal{F}\left[H\left(x,y\right)e^{j |\textbf{k}| (x^2+y^2)/2d}\right]$, where $\mathcal{F}$ is the FFT operator, and $d$ is the propagation distance to refocus over the pupil (Fig. \ref{fig_BW}(b)). The pupil image is then cropped and recentered leading to $H_2(k_x,k_y)$ (Fig. \ref{fig_BW}(c)) and the complex object field $H_3\left(x,y,z\right)$ is extracted by  $H_3\left(x,y,z\right)=\mathcal{F}^{-1}\left[H_2(k_x,k_y)~e^{j(k_x^2+k_y^2)/2z}\right]$.

%\section{Fully-quantitative laser Doppler imaging}

Due to the red blood cell (RBC) motion, the light passing through the zebrafish undergoes a Doppler shift $\omega_{\rm D}=(\textbf{k}_{\rm S}-\textbf{k}_{\rm I})\textbf{.}\textbf{v}$, where   $\textbf{k}_{\rm I}$ and $\textbf{k}_{\rm S}$ are  the illumination and scattered wavevectors, and  $\textbf{v}$ is the velocity.
To measure the Doppler shift due to scatterer motion, we have swept the detection frequency $\Delta\omega$ from 0 to 120 Hz, and calculated two-phases   holograms $H=I_0-I_1$ for each frequency point.
\begin{figure}[htbp]
\centering
\includegraphics[height = 6.0cm]{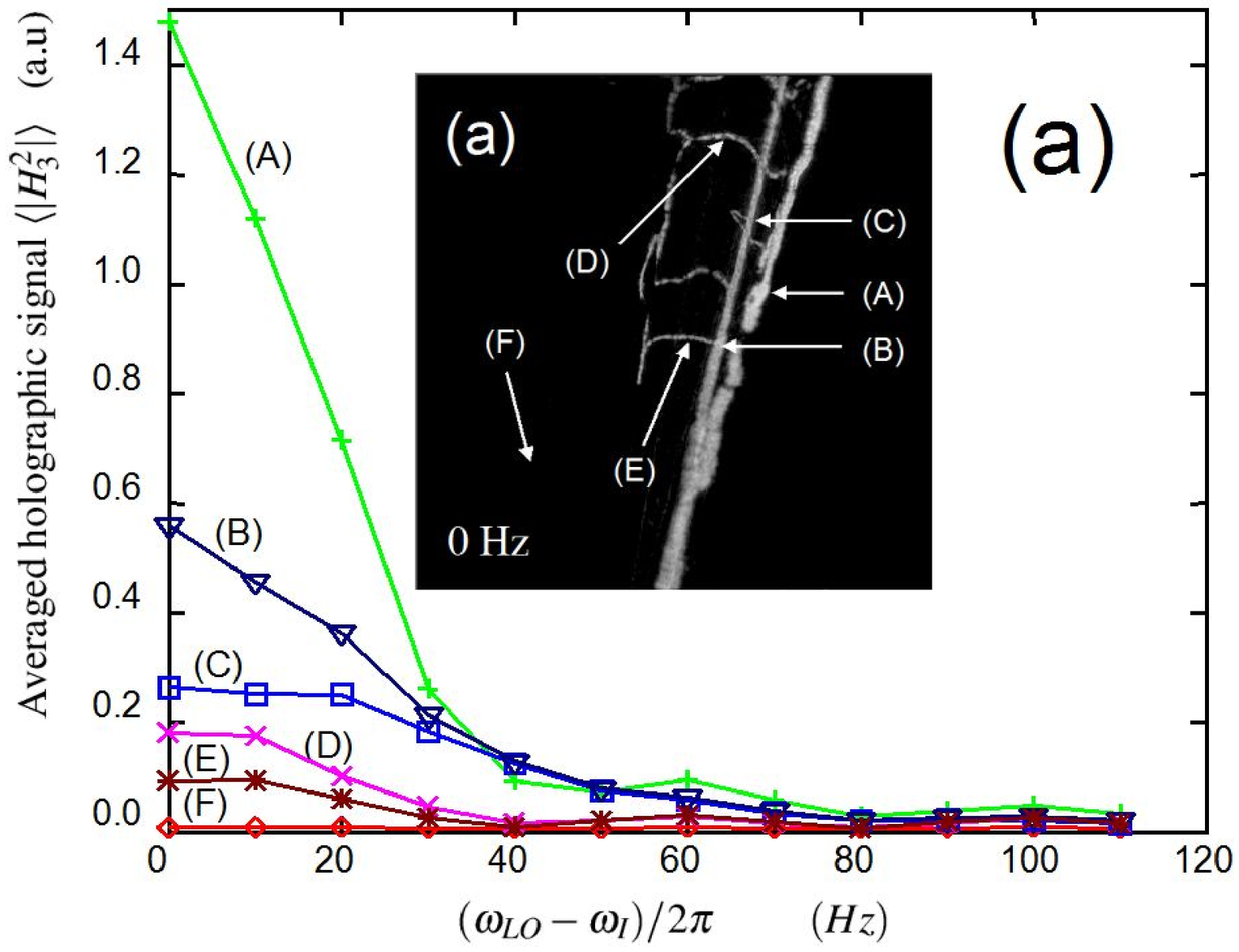}
\includegraphics[height = 6.0cm]{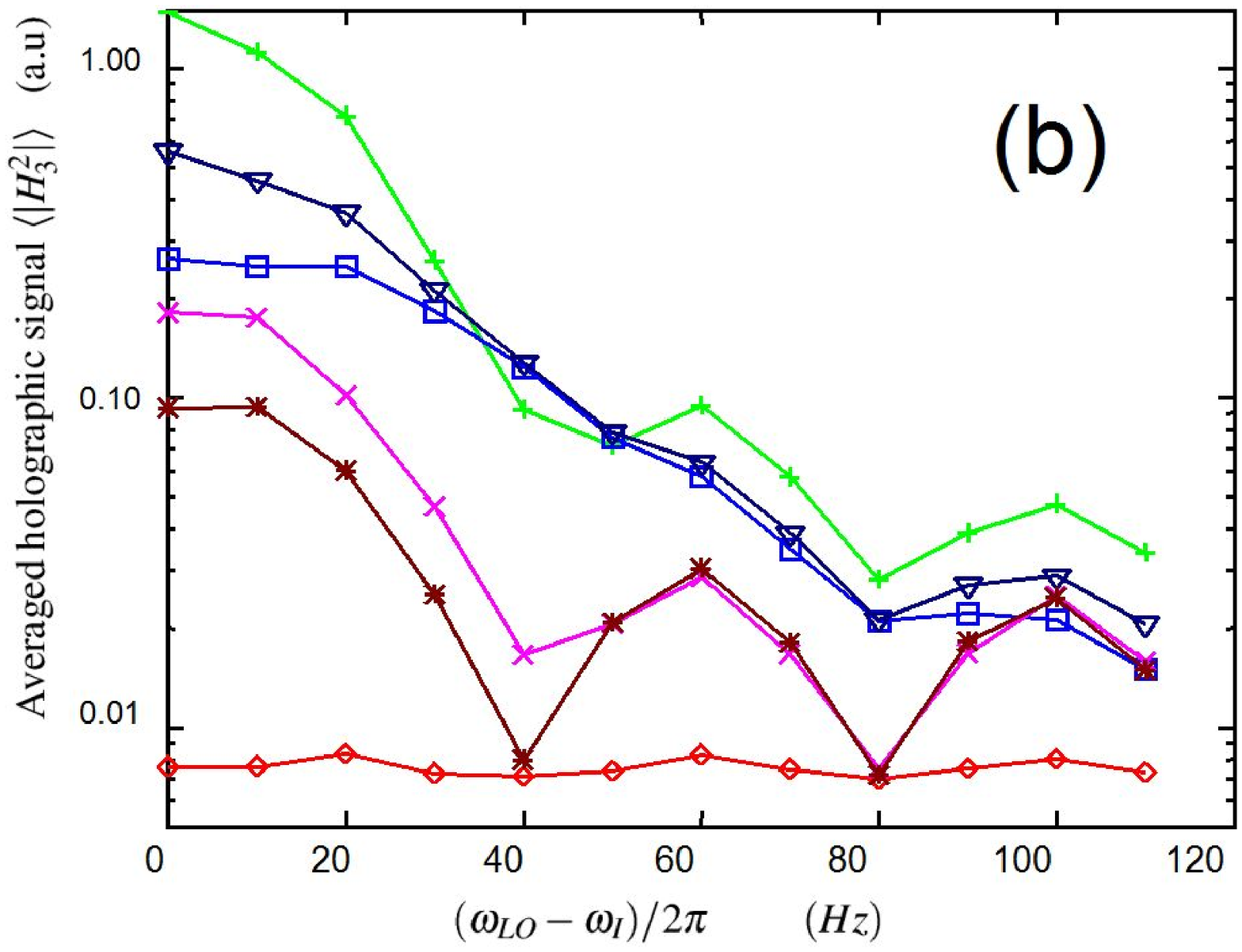}
\caption{Dependance of the Doppler holographic signal $\langle |H_3(x,y)|^2\rangle $ with the frequency offset $\Delta \omega=(\omega_{LO}-\omega_{\rm I})/(2\pi)$ for different location A to F  in the insert (a). Curves are drawn with linear (a) and logarithmic (b) scales.}
\label{fig_curve}
\end{figure}
To assess for the norm of the velocity vector, the signal is averaged over different zones the reconstructed image. Obtained results are depicted Fig. \ref{fig_curve} for zones labeled in the insert of Fig. \ref{fig_curve}(a).
By estimating the half-width at half-maximum of each spectra, it is possible to obtain the norm of the velocity vector.
\begin{figure}[htbp]
\centering
\includegraphics[width = 5.2cm]{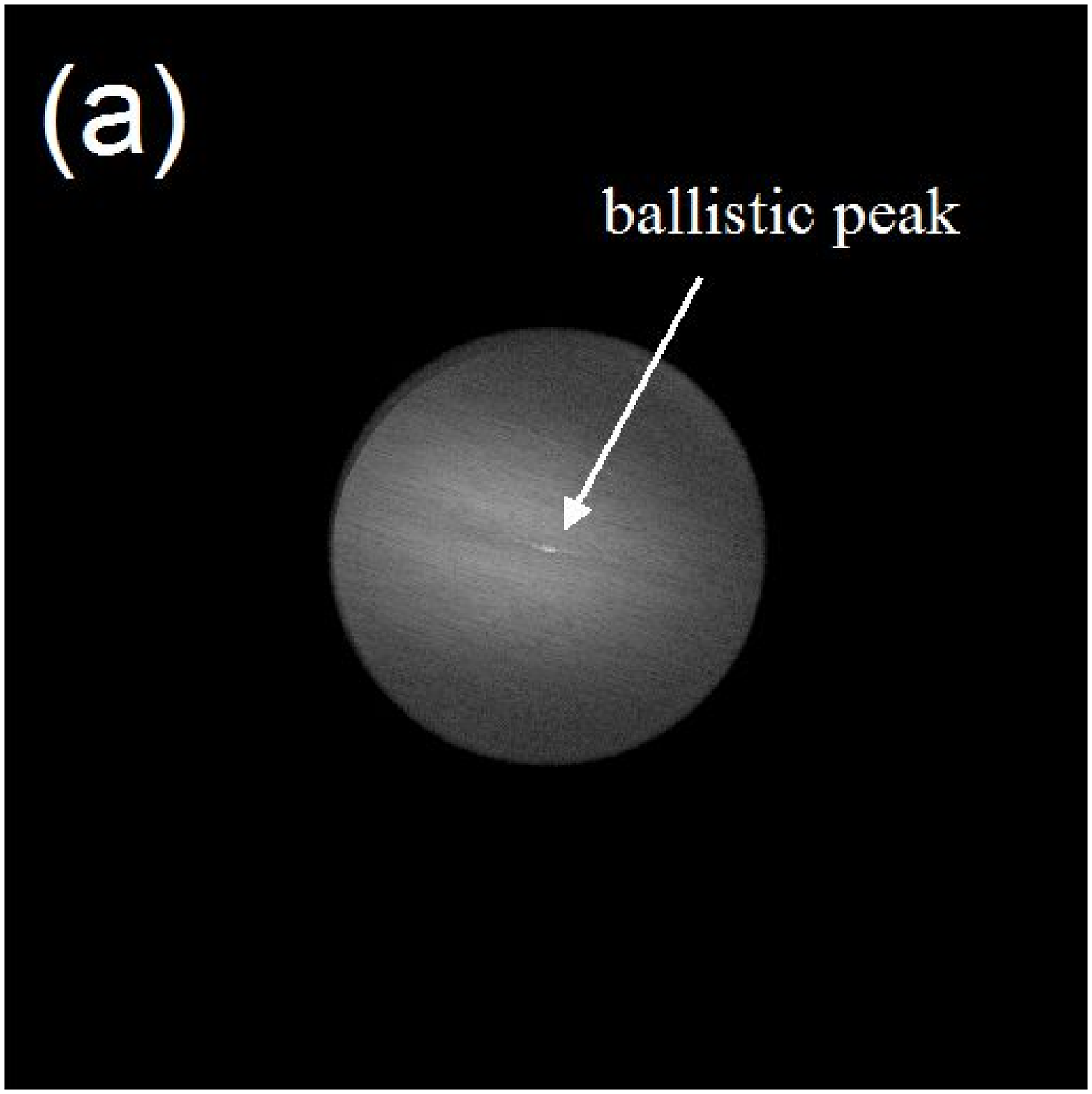}
\includegraphics[width = 5.2cm]{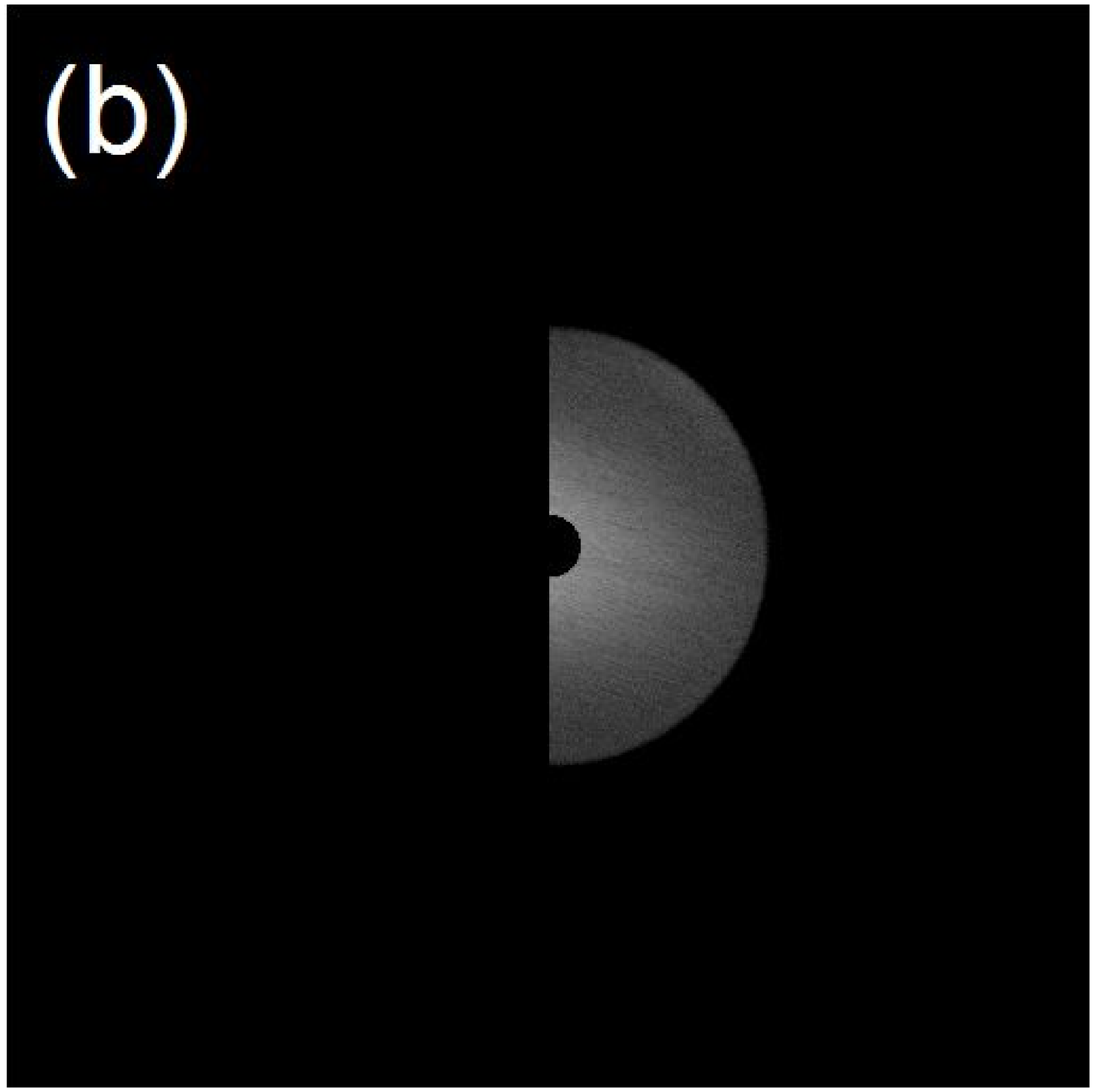}
\includegraphics[width = 5.2cm]{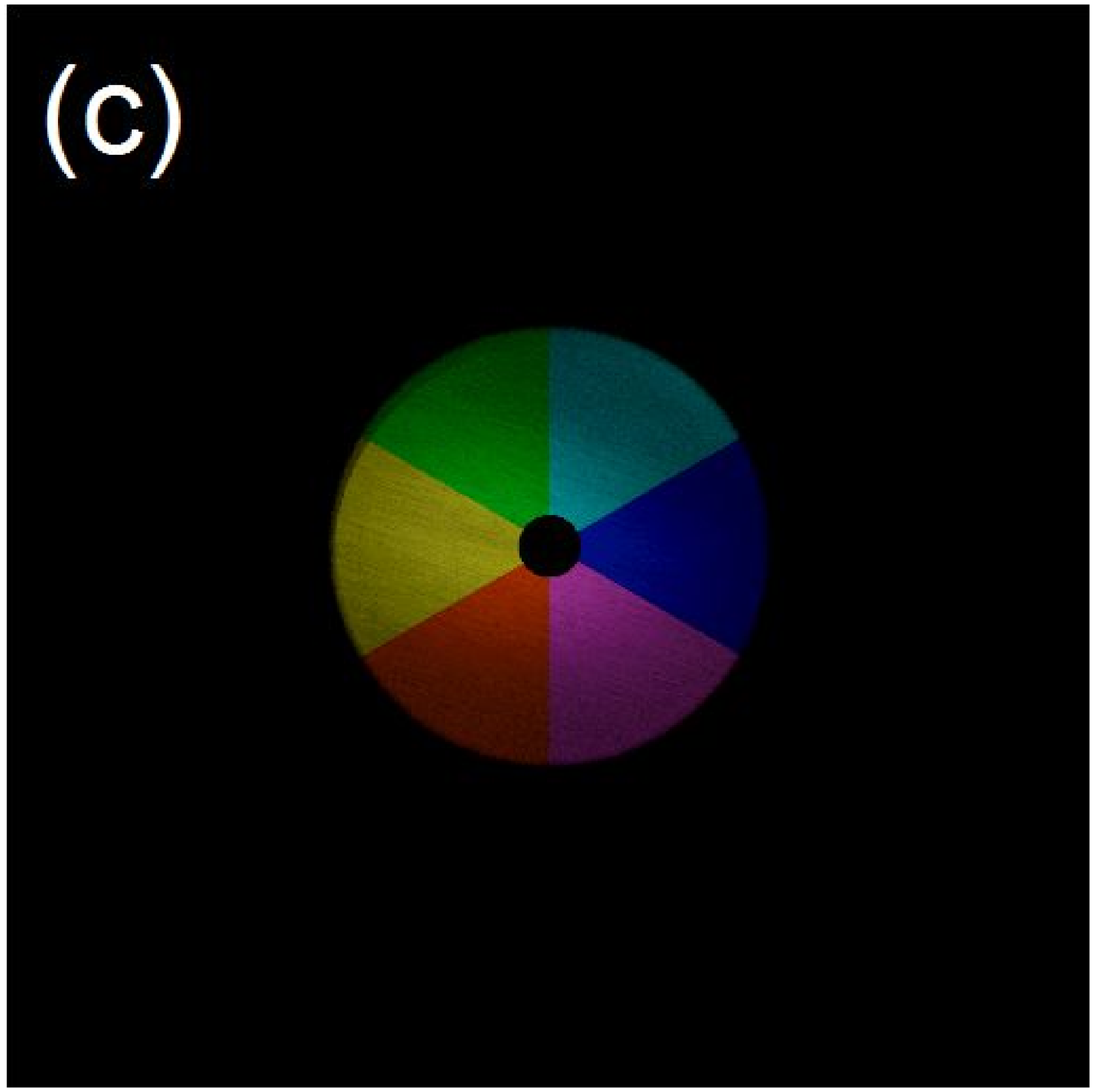}
\caption{Fourier space reconstructed hologram  $H_2(k_x,k_y)$ made without (a) and with (b,c) selection of the scattered wave vector $\textbf{k}_S$. In (b) the selected zone is oriented in $x$ direction. In (c), three zones obtained by rotating (b)  by 0 (blue),  $2\pi/3$ (green) and  $4\pi/3$ (red) are displayed.   Display is made in arbitrary log scale for  $\langle |H_{2}|^2\rangle$. Frequency shift is $(\omega_{LO}-\omega_{\rm I})=0$ . }
\label{fig_5k}
\end{figure}

\begin{figure}[htbp]
\centering
\includegraphics[width = 5.2cm]{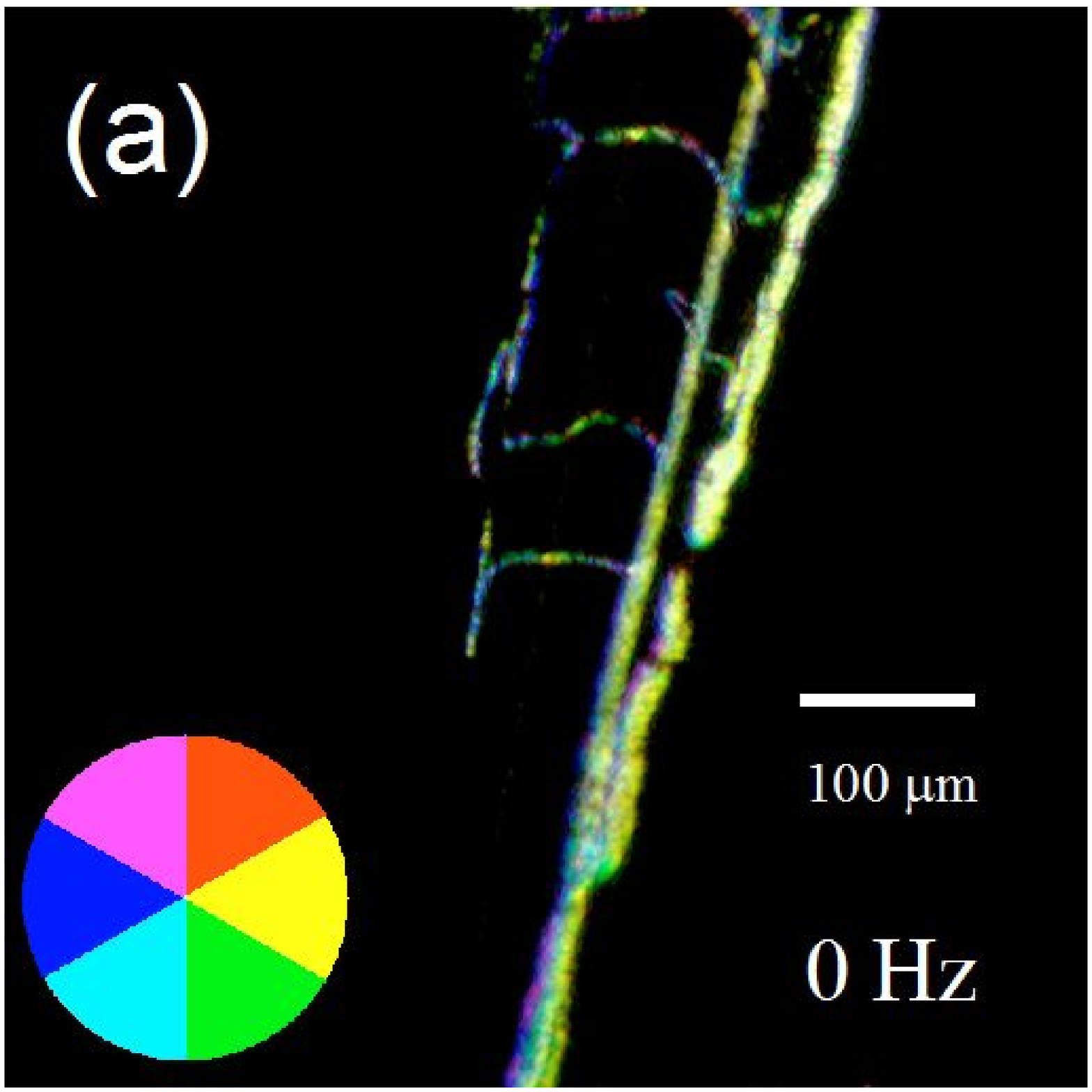}
\includegraphics[width = 5.2cm]{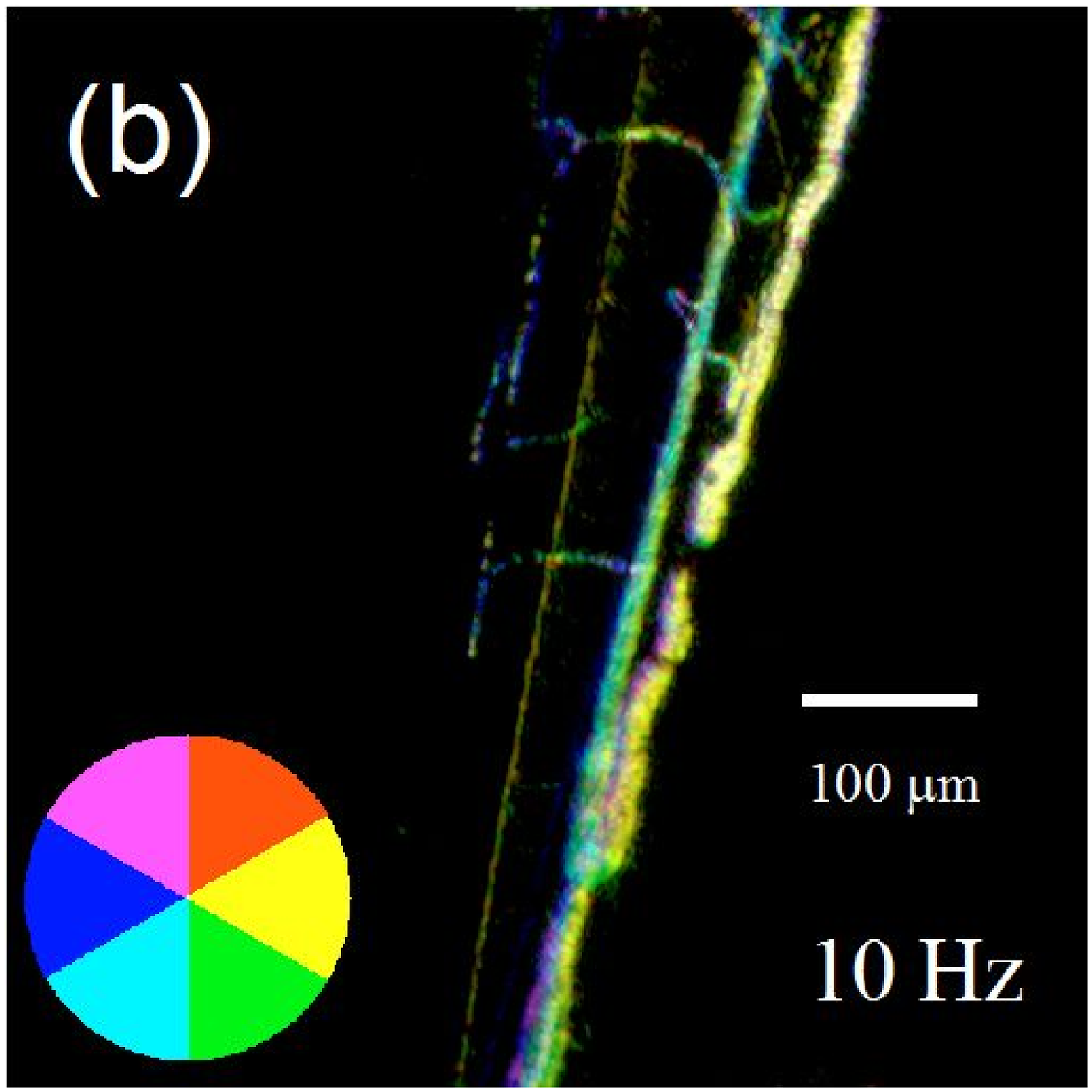}
\includegraphics[width = 5.2cm]{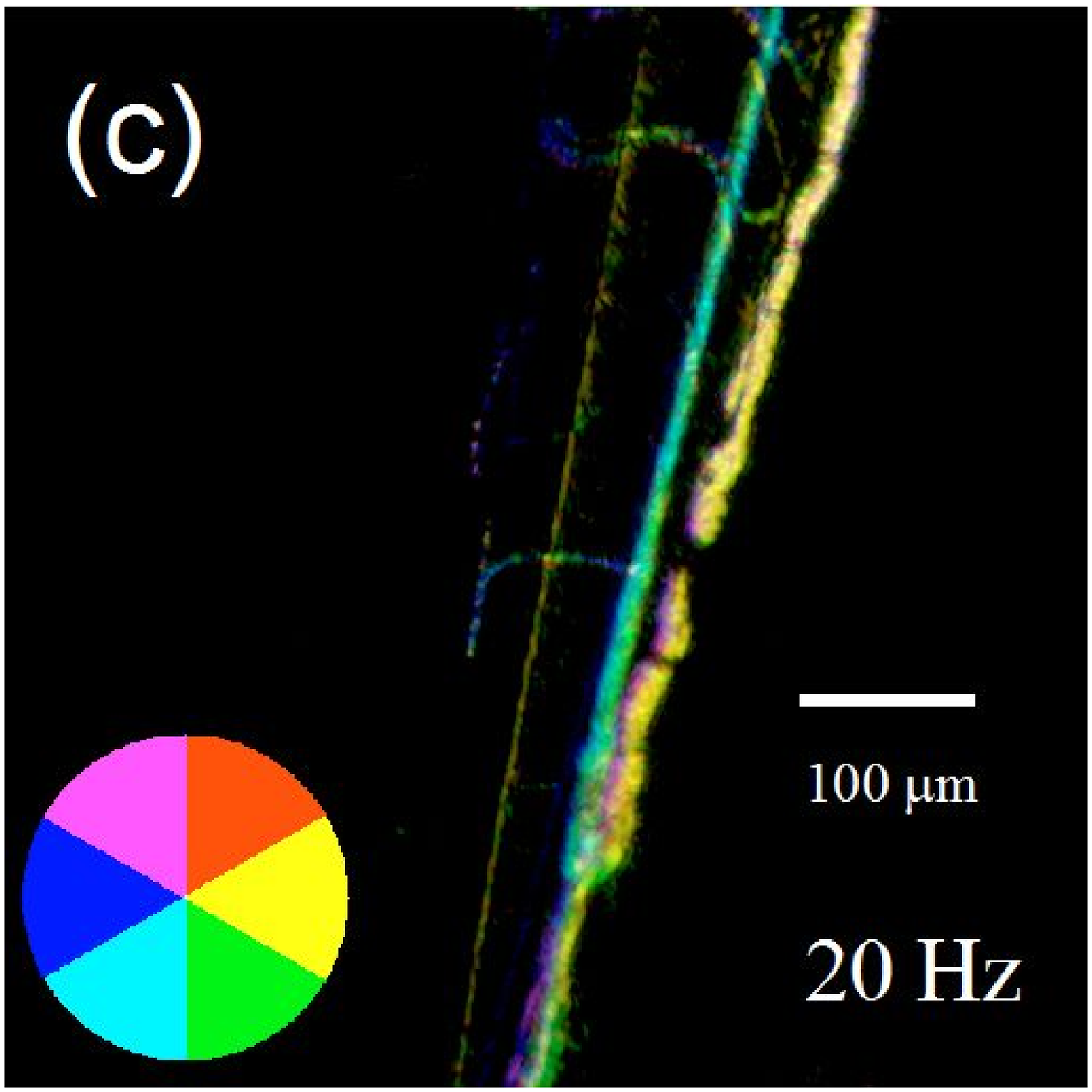}
\caption{Colored reconstructed hologram  $H_3(x,y,z=0)$ made for $(\omega_{LO}-\omega_I)/(2\pi)=
 0 $ Hz (a), 10 Hz (b), 20 Hz (c) displayed with a color RGB Log scale for the average intensity  $\langle |H_{3}|^2\rangle$. }
\label{fig_5}
\end{figure}

In our experiment, the reference arm frequency offset $\Delta \omega$ (which is known)  is close to the Doppler shift: $(\textbf{k}_S-\textbf{k}_I)\textbf{.} \textbf{v}$, where $\textbf{k}_{\rm I}$  is know ($\textbf{k}_{\rm I}$ is parallel to the $z$ axis), and where $\textbf{k}_{\rm S}$ can be  selected in the Fourier space (by applying  mask to $H_2(k_x,k_y)$)). Quantitative information on the velocity $\textbf{v}$ can thus be obtained.  This point is illustrated by Fig. \ref{fig_5k} and Fig \ref{fig_5}. By filtering half of the Fourier space (Fig. \ref{fig_5k}(b)), one can discriminate RBC motion along the $x$ direction. By rotating the filtering zone  by 0 (blue), $2\pi/3$ (green) and  $4\pi/3$ (red), one can map in false color the reconstruction space (see Fig. \ref{fig_5k}(c)).
Combining the 3 reconstructed  holograms (displayed in blue, green and red) for the 3 rotations of the filtering zone, one can obtain an image  where the flow-direction is coded in RGB colors. This point is illustrated Fig. \ref{fig_5} where blood flow direction between veins and arteries can be discriminated.

\section{Conclusion}
We have proposed an holographic laser Doppler set-up based on heterodyne digital holography working in transmission configuration, and fitted to an upright-microscope. Upon classical holographic modalities, coupling spectral broadening measurement with RGB fluid flow-direction discrimination has made possible to obtain a quantitative Doppler measurement.

We acknowledge OSEO-ISI Datadiag grant, ANR Blanc Simi 10 (n. 11 BS10 015 02) grant, and Labex Numev (convention ANR-10-LABX-20) n. 2014-1-042 grant
for funding.

\end{document}